\documentclass[conference]{IEEEtran}
\usepackage{graphicx}
\usepackage{amsmath}
\usepackage{amssymb}
\usepackage{array}
\usepackage{cite}
\begin{document}
%
\title{Overlapped-MIMO Radar Waveform Design for Coexistence With Communication Systems}
%
\author{\IEEEauthorblockN{Chowdhury Shahriar, Ahmed Abdelhadi and T. Charles Clancy}
\IEEEauthorblockA{Ted and Karyn Hume Center for National Security and Technology\\
Bradley Department of Electrical and Computer Engineering\\
Virginia Tech, Arlington, VA 22203, USA\\
Email: \{cshahria, aabdelhadi, tcc\}@vt.edu}}
\maketitle

\footnotetext[1]{{\bf This work was sponsored by the Defense Advanced Research Projects Agency (DARPA) under the Shared Spectrum Access for Radar and Communications (SSPARC) program Contract DARPA-BAA-13-24. The views expressed are those of the authors and do not reflect the official policy or position of the Department of Defense or the U.S. Government. Distribution Statement A: Approved for Public Release, Distribution Unlimited.}}

\begin{abstract}
This paper explores an overlapped-multiple-input multiple-output (MIMO) antenna architecture and a spectrum sharing algorithm via null space projection (NSP) for radar-communications coexistence. In the overlapped-MIMO architecture, the transmit array of a collocated MIMO radar is partitioned into a number of subarrays that are allowed to overlap. Each of the antenna elements in these subarrays have signals orthogonal to each other and to the elements of the other subarrays. The proposed architecture not only improves sidelobe suppression to reduce interference to communications system, but also enjoys the advantages of MIMO radar without sacrificing the main desirable characteristics. The radar-centric spectrum sharing algorithm then projects the radar signal onto the null space of the communications system's interference channel, which helps to avoid interference from the radar. Numerical results are presented which show the performance of the proposed waveform design algorithm in terms of overall 
beampattern and sidelobe levels of the radar waveform and finally shows a comparison of the proposed system with existing collocated MIMO radar architectures.
\end{abstract}

\IEEEpeerreviewmaketitle

\section{Introduction}
Modern wireless communication systems have evolved towards higher throughput within limited bandwidth in order to accommodate the growing demand for high-volume data streams, such as video broadcasting. The tremendous growth of data services via broadband wireless access (BWA) has resulted in scarcity of wireless bandwidth in recent years. This has prompted government entities like the Federal Communications Commission (FCC) and National Telecommunications and Information Administration (NTIA) to explore options for new bandwidth and reassign underused spectrum. They are interested in sharing spectrum previously used and reallocating some of the bandwidth that had been assigned to Department of Defense (DoD). Recently in its $2010$ Fast Track Report, NTIA proposed to share the $3550-3650$ MHz band between military radars and commercial BWA communication systems~\cite{NTIA10}. According to the NTIA report, this band is under-utilized and is favorable for BWA standards such as LTE to coexist with radars.
 
However, coexistence will require mitigation of electromagnetic interferences (EMI) from one to another. In this case, it will be mostly dominated by radar EMI. Cellular wireless devices and base stations transmit on the order of milliwatts and microwatts, whereas radars transmit up to megawatts. Quite naturally we are here focusing on designing radar waveforms to avoid EMI to communications systems. 

\label{sec:introduction}
\subsection{Related Work}

The concepts of MIMO radar are getting attention nowadays as they can have better performance than legacy radar systems to identify more targets with higher angular resolution~\cite{Li2009}. In MIMO radar, multiple waveforms are transmitted via multiple transmit antenna elements and reflected signals from the targets are received by multiple receive antennas. In~\cite{Hassanien2010}, authors have proposed a different kind of MIMO radar that they called Phased-MIMO radar. In Phased-MIMO radar, waveforms are transmitted from a MIMO radar where antenna elements are partitioned into multiple subarrays and the elements are allowed to overlap among subarrays. The benefit of this formulation over conventional MIMO radar is its higher coherent processing gain and overall suppressed sidelobes. 

The idea of projecting signals onto the null space of an interference channel, in order to avoid interference, is a well-studied topic in the cognitive radio research community~\cite{YiICC2010, NoamICC2012}. An interference channel's null space is calculated at the transmitter either by exploiting channel reciprocity using its second order statistics~\cite{YiICC2010} or by blindly estimating the null space, if no cooperation exists between resource sharing nodes~\cite{NoamICC2012}. However, for MIMO radar systems this idea of null space projection (NSP) was first proposed in~\cite{Shabnam2012GC}, which was followed by an array of papers ~\cite{KhawarICNC14, Awais_Spatial, KAC14ICC} where authors studied the NSP-based spectrum sharing approach for various radar-communications scenarios to avoid interference on communications system. 

\subsection{Our Contributions}
In this paper, we extend the previous work of radar signal projection onto the null space of the interference channel between radar and communication systems in order to avoid interference to communications systems~\cite{Shabnam2012GC,KhawarICNC14, Awais_Spatial}. The previous work considers a coexistence scenario where MIMO radar operates in the same band as a MIMO communications system. We extend this approach and consider a different MIMO radar formulation that we call the overlapped-MIMO radar. In the overlapped-MIMO architecture, the transmit array of a collocated MIMO radar is partitioned into a number of subarrays that are allowed to overlap. Each of the antenna elements in these subarrays have signals orthogonal to each other and to the elements of the other subarrays. The proposed architecture not only provides better sidelobe suppression to reduce the interference to communications systems, but also enjoys the advantages of MIMO radar without sacrificing the main desirable characteristics. The 
radar-centric spectrum sharing algorithm is then combined with this overlapped-MIMO radar formulation that enables the radar signal to project onto the null space of the interference channel of the communications system.

Our contributions in this paper are as follows:

\begin{itemize}
	\item We introduce a new formulation of MIMO radar, called overlapped-MIMO radar, where the transmit array of the radar is partitioned into a number of subarrays that are allowed to overlap. We derive an analytical model of the architecture to establish the validity of our proposal. Through numerical results and simulation, we show that this architecture results in better sidelobe suppression than conventional radar, which improves the coexistence between radar and communications systems. 
	\item We extend the spectrum sharing scenario of~\cite{Shabnam2012GC} between a MIMO radar and a communications system to spectrum sharing between an overlapped-MIMO radar and a communications system. This extension gives rise to a different coexistence scenario as the overall beampattern of this radar waveform is significantly different from conventional MIMO radar. 
\end{itemize}

The remainder of this paper is organized as follows. Section II builds the foundation of spectrum sharing architecture between MIMO radar and MIMO communications system. Section III discusses the preliminaries of collocated MIMO radar. Section IV introduces the overlapped-MIMO radar. Section V presents the NSP projection algorithm. Section VI discusses the simulation setup and provides quantitative results along with discussion. Section VII concludes the paper.

\section{Spectrum Sharing Architecture}
\label{sec:spectrum}

In this paper, we consider a scenario where a collocated MIMO radar with $M_T$ transmit and $M_R$ receive antennas coexists in the same band with a MIMO communications system with $N_T$ transmit and $N_R$ receive antennas. 

In this case, the received signal at the receiver terminal of the communications system can be written as

\begin{equation}
\label{S2EQ1}
\mathbf{y_{C}}(t) = \mathbf{H_{I}}^{N_R \times M_T} \mathbf{x_{Radar}} (t) + \mathbf{H}^{N_R \times N_T} \mathbf{x_{C}} (t) + \mathbf{n}(t)
\end{equation}

\noindent where $\mathbf{x_{Radar}} (t)$ is transmitted radar signal, $\mathbf{x_{C}} (t)$ is transmitted communications signal, $\mathbf{H_I}$ is $N_R \times M_T$ interference channel between radar and communications system, $\mathbf{H}$ is $N_R \times N_T$ channel between transmitter and receiver of the communications system, $\mathbf{n}(t)$ is additive white Gaussian noise (AWGN). 

We assume that both the radar and communications system are working in a friendly environment, cooperating with each other and sharing information. Information is shared under an agreement that each system will seek to avoid causing interference to the other. 

\begin{figure}
\centering
\includegraphics[width=3.2in]{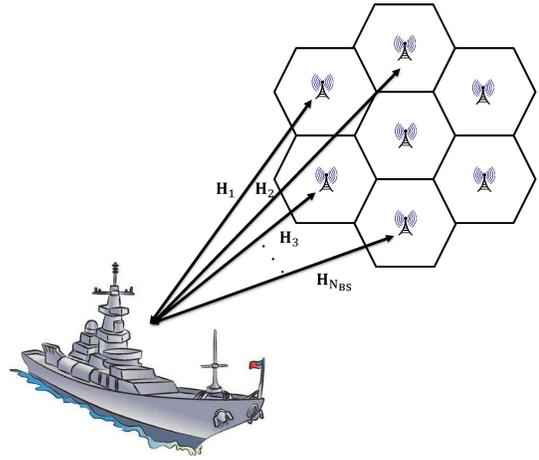}
\caption{A possible spectrum sharing scenario between a radar mounted on a ship and an on shore communications system.}
\label{fig:spectrum_share}
\end{figure}

In this paper we investigate a radar-centric design approach. In our radar-centric design, we assume that the ICSI of the communications system is available at the radar terminal and the goal of the MIMO radar is to develop its own radar waveform that will avoid interference to the communications system. A typical coexistence scenario is shown in Fig.~\ref{fig:spectrum_share}. 

\section{Collocated MIMO Radar}
\label{sec:MIMO}

In this section, we derive the preliminaries of the collocated MIMO radar system. The MIMO radar considered in this paper is collocated, which means transmit and receive antennas are assumed to be close to each other in space (possibly the same array)~\cite{Li2009}. The number of antenna elements in transmit and receive array are assumed as $M_T$ and $M_R$. 

Let $\boldsymbol{\phi}(t)$ be the waveform emitted from the MIMO radar, which is defined as 

\begin{equation}
\label{S3EQ1}
\boldsymbol{\phi} (t) = 
\begin{bmatrix} 
{\phi}_1(t) & {\phi}_2(t) & \cdots & {\phi}_{M_T}(t) 
\end{bmatrix}^{T} 
\end{equation}

\noindent where $t$ is the time index within the radar pulse and $(\cdot)^T$ denotes the transpose. Note that the $m$th transmit antenna emits the $m$th element of the vector $\boldsymbol{\phi} (t)$, which is $\phi_{m} (t)$. We assume that each element of the transmitted waveform is orthogonal to each other, and the overall waveform satisfies the orthogonality principle

\begin{equation}
\label{S3EQ2}
\int_{T_0} \boldsymbol{\phi} (t) \boldsymbol{\phi}^{H} (t) dt = \mathbf{I}_{M_T}
\end{equation}

\noindent where $T_0$ is the radar pulse width, $(\cdot)^H$ stands for the Hermitian transpose and $\mathbf{I}_{M_T}$ is the $M_T \times M_T$ identity matrix. 

The transmitter waveform is steered towards a specific target (or source) direction. Assuming $\mathbf{a}(\theta)$ is the actual transmit steering vector of size $M_T \times 1$ associated with the direction $\theta$ for a uniform linear array (ULA), then

\begin{eqnarray}
\label{S3EQ3}
\mathbf{a} (\theta) 	&=& \begin{bmatrix} a_1(\theta) & a_2(\theta) & \cdots &  a_{M_T}(\theta) \end{bmatrix}^{T} \\ \notag
											&=& \begin{bmatrix} 1 & e^{-j 2 \pi d_T \sin\theta} & \cdots & e^{-j 2 \pi d_T (M_T - 1) \sin\theta} \end{bmatrix}^{T} \notag									
\end{eqnarray}

\noindent where the first element of $\mathbf{a}(\theta)$ is the reference element $a_{1} (\theta) = 1$, $m$th element is $a_{m} (\theta) = e^{-j 2 \pi d_T m \sin\theta}$, and $d_{T}$ is the inter-element spacing measured in wavelength for the array. So, the waveform is steered with the steering vector and finally the transmitted signal vector can now be written in a compact vector form as 
%

\begin{eqnarray}
\label{S3EQ4}
\mathbf{x_{Radar}} (t) &=& \mathbf{a} (\theta) \odot \boldsymbol{\phi} (t)  \\ \notag
											 &=& \begin{bmatrix} a_1(\theta){\phi}_1(t) & a_2(\theta){\phi}_2(t) & \cdots & a_{M_T}(\theta){\phi}_{M_T}(t)  \end{bmatrix} \\ \notag
											 &=& \begin{bmatrix} x_1(t) & x_2(t) & \cdots & x_{M_T}(t) \end{bmatrix} \notag 
\end{eqnarray}

\noindent where $\odot$ is the Hadamard (element-wise) product.

The $M_R \times 1$ snapshot vector received by the receive antenna array can be modeled as 

\begin{equation}
\label{S3EQ5}
\mathbf{y_{Radar}} (t) = \mathbf{y_{s}} (t) + \mathbf{y_{i}} (t) + \mathbf{n} (t)
\end{equation}

\noindent where $\mathbf{y_{s}} (t)$ is the target/source signal, $\mathbf{y_{i}} (t)$ is a jamming/interference signal, and $\mathbf{n} (t)$ is AWGN.

Under the single point target/source assumption, the received signal at the radar can be written as

\begin{equation}
\label{S3EQ6}
\mathbf{y_{s}} (t) = \beta_{s} ( \mathbf{a}^{T}(\theta_{s})\boldsymbol{\phi}(t) ) \mathbf{b}(\theta_{s})
\end{equation}

\noindent where $\theta_{s}$ is the target/source direction, $\beta_{s}$ is complex-valued reflection coefficient of the focal point $\theta_{s}$ (that includes channel effect and propagation loss), and $\mathbf{b}(\theta)$ is the $M_R \times 1$ receive steering vector associated with the direction $\theta$.

The returned signal from $m$th transmitted waveform can be found by matched-filtering the received signal to each of the waveforms $\{\phi_{m} (t)\}^{M_T}_{m=1}$, i.e.,

\begin{equation}
\label{S3EQ7}
\mathbf{y_{m}} (t) = \int_{T_0} \mathbf{y_{Radar}} (t) {\phi}^{*}(t)  dt \quad m=1,\cdots,M_T
\end{equation}

\noindent where $(\cdot)$ stands for the conjugate operation. Then, the virtual data vector of size $M_T M_R \times 1$ can be written as

\begin{eqnarray}
\label{S3EQ8}
\mathbf{y_{v}} &=& \left[ \mathbf{y}_{1}^T \mathbf{y}_{2}^T \cdots \mathbf{y}_{M_T}^T \right]^T \\ \notag
               &=& \beta_{s} \mathbf{a} (\theta_{s}) \otimes \mathbf{b} (\theta_{s}) + \mathbf{y_{i+n}}  \notag
\end{eqnarray}

\noindent where $\otimes$ is the Kronker product operator and $\mathbf{y_{i+n}}$ is combined component of interference and noise. Hence, the target/source signal component can be written as 

\begin{equation}
\label{S3EQ9}
\mathbf{y_{s}} = \beta_{s} \mathbf{v} (\theta_{s}) 
\end{equation}

\noindent where $\mathbf{v} = \mathbf{a} (\theta_{s}) \otimes \mathbf{b} (\theta_{s}) $ is the $M_T  M_R \times 1$ virtual steering vector associated with a virtual array of $M_T M_R$ elements.
 
For a ULA, the $(m_t M_R + m_r)$th entry of the virtual array steering vector $\mathbf{v }(\theta )$ can be expressed as

\begin{equation}
\label{S3EQ10}
\mathbf{v}_{\left[ m_t M_R + m_r \right]} (\theta) = e^{-j 2 \pi ( m_t d_T \sin\theta + m_r d_R \sin\theta )}                                             
\end{equation}

\noindent where $m_t = 0, \cdots, M_T-1$ and $m_r = 0, \cdots, M_R-1$. For $d_T = M_R d_R$, the virtual array steering vector simplifies to~\cite{Chen2008} 

\begin{equation}
\label{S3EQ11}
\mathbf{v}_{\left[ \varsigma \right]} (\theta) = e^{-j 2 \pi \varsigma d_R \sin\theta }                              
\end{equation}

\noindent where $\varsigma = m_t M_R + m_r = 0,1, \cdots, M_T M_R - 1$, which means that an $M_T M_R$ effective aperture array can be obtained by using $M_T + M_R$ antennas~\cite{Hassanien2010}. Note that in this case, the resulting virtual array is a ULA of $M_T M_R$ elements spaced $d_R$ wavelength apart from each other.

\section{Proposed Overlapped-MIMO Radar}
\label{sec:O-MIMO}

\begin{figure}
\centering
\includegraphics[width=3.2in]{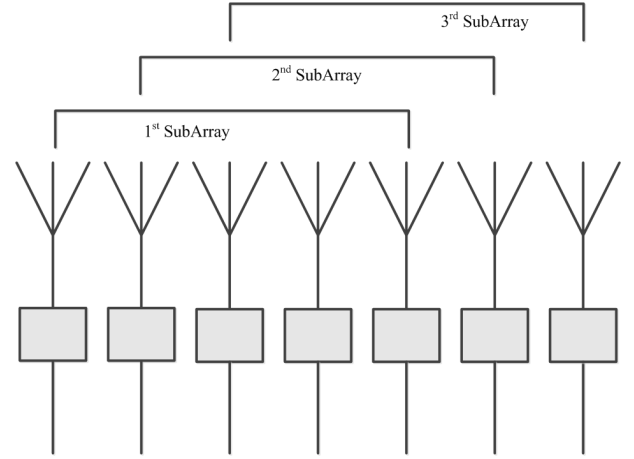}
\caption{A block diagram of the overlapped-MIMO radar formulation.}
\label{fig:omimoblock}
\end{figure}

In this section, we introduce a new formulation of MIMO antenna arrays called overlapped-MIMO where we partition the array into multiple overlapped subarrays. One of the advantages of this approach is that it allows us to beamform transmit and receive arrays. The key idea behind this approach is to partition the transmit arrays into $K$ subarrays where $1\leq K \leq M_T$, which are allowed to overlap~\cite{Hassanien2010}. The overlapped-MIMO radar formulation is shown in Fig.~\ref{fig:omimoblock}. 

The complex envelope of the signal at the output of the $k$th subarray can be expressed as

\begin{equation}
\label{S4EQ1}
s_k = \sqrt{\frac{M_T}{K}} \boldsymbol{\phi}_{k} (t) \odot \boldsymbol{w}_{k} \quad k = 1,\cdots, K                         
\end{equation}

\noindent where both $\boldsymbol{\phi}_{k}$ and $\boldsymbol{w}_{k}$ are $M_m \times 1$ vectors. The latter one has a unit-norm complex vector with $M_m$ beamforming weights. The former one is a waveform vector with $M_m$ orthogonal waveforms. Note that here $M_m$ is the number of antenna elements is each subarray and is defined as $M_m = M_{T} - K + 1$. Each of the orthogonal waveforms, $\phi(t)^{m}_{k}$ is a transmitted signal which can be modeled as 

\begin{equation}
\phi^{m}_{k} = Q(t) e^{j 2 \pi (mk/T_0)t}
\label{eq:phmimo02}
\end{equation}

\noindent where $Q(t)$ is the pulse shape of duration $T_{0}$, $m = 1, \cdots , M_m$ and $k = 1, \cdots , K$~\cite{He2010}. 

The reflected signal from a target located at direction $\theta$ in the far-field can be expressed as

\begin{equation}
r(t,\boldsymbol{\theta}) = \sqrt{\frac{M_T}{K}} \beta(\boldsymbol{\theta}) \displaystyle \sum^{K}_{k=1} \sum^{M_m}_{m=1} w^{m}_{k} {\phi}^{m}_{k} (t) d^{m}_{k}(\boldsymbol{\theta})
\end{equation}

\noindent where $\beta(\boldsymbol{\theta})$ is the reflection coefficient (constant), and $d^{m}_{k}(\boldsymbol{\theta})$ is the waveform diversity vector, which is defined as $\displaystyle{e^{-j \tau^{m}_{k}(\boldsymbol{\theta})}}$ where $\tau^{m}_{k}(\boldsymbol{\theta})$ is the time required for the wave to travel from the first element to the next element. 

The received complex vector of the array observation can be written as

\begin{equation}
\mathbf{y_{Radar}} (t) = r(t,\boldsymbol{\theta}_s) \mathbf{b}(\boldsymbol{\theta}_s) + \sum^{D}_{i} r(t,\boldsymbol{\theta}_i) \mathbf{b}(\boldsymbol{\theta}_i) + \boldsymbol{n}(t)
\end{equation}

\noindent where $D$ is the number of interfering signals, $\mathbf{b}(\boldsymbol{\theta})$ is the receive steering vector of size $M_R \times 1$ associated with direction $\boldsymbol{\theta}$, and $\boldsymbol{n}(t)$ is AWGN. 

Following equations (\ref{S3EQ7}) and (\ref{S3EQ8}), by match-filtering $ \mathbf{y_{Radar}} (t)$ to each of the waveforms $\left\{\phi^{m}_{k}\right\}^{M_m,K}_{m=1,k=1}$ we can obtain $M_m K M_R \times 1$ virtual data vectors as
 
\begin{equation}
\mathbf{y_{v}} = \sqrt{\frac{M_T}{K}} \beta_{s} \mathbf{u} (\theta_{s}) + \sum^{D}_{i} \sqrt{\frac{M_T}{K}} \beta_{i} \mathbf{u} (\theta_{i}) + \boldsymbol{n}(t)
\end{equation}

\noindent where $\mathbf{u} (\theta) = (\mathbf{c} (\theta) \odot \mathbf{d} (\theta)) \otimes \mathbf{b} (\theta)$ is the $M_m K M_R \times 1$ virtual steering vector, intermediate vector $c=\left\{w^{m}_{k}a^{m}_{k}\right\}^{M_m,K}_{m=1,k=1}$ of size $M_m K \times 1$, and waveform diversity vector $d=\left\{\exp{{-j \tau^{m}_{k}(\boldsymbol{\theta})}}\right\}^{M_m,K}_{m=1,k=1}$ of size $M_m K \times 1$~\cite{Chen2008}.

In the case of non-adaptive beamforming, the corresponding beamformer weight vectors are given for the $k$th transmitting subarray as

\begin{equation}
\mathbf{w}_k = \frac{\mathbf{a}_k (\theta_{s})}{\left\|\mathbf{a}_k (\theta_{s})\right\|} = \frac{\mathbf{a}_k (\theta_{s})}{\sqrt{M_T-K+1}}        
\end{equation}

\noindent where $k = 1,2, \cdots, K$. They are given for the receiving subarrays as

\begin{equation}
\mathbf{w}_d = (\mathbf{c}(\boldsymbol{\theta}_s) \odot \mathbf{d}(\boldsymbol{\theta}_s)) \otimes  \mathbf{b}(\boldsymbol{\theta}_s)   
\end{equation}

Let $G(\theta)$ be the normalized overall beampattern 

\begin{equation}
\label{EQ:BP}
G(\theta) = \frac{| \mathbf{w}^{H}_d \mathbf{u}(\boldsymbol\theta)|^{2}}{| \mathbf{w}^{H}_d \mathbf{u}(\boldsymbol\theta_s)|^{2}} = \frac{| \mathbf{u}^H(\boldsymbol\theta_s) \mathbf{u}(\boldsymbol\theta)|^2}{\left\| \mathbf{u}(\boldsymbol\theta_s)\right\|^4}
\end{equation}

For the special case of a ULA, we have $\mathbf{a}^{H}_{1}(\boldsymbol\theta) \mathbf{a}_{1} (\boldsymbol\theta_s) = \cdots = \mathbf{a}^{H}_{K}(\boldsymbol\theta) \mathbf{a}_{K} (\boldsymbol\theta_s)$. Using equation (\ref{EQ:BP}), the beampattern of the overlapped-MIMO radar for a ULA with overlapped partitioning of $K$ transmit subarrays can expressed as 

\begin{equation}
G_K(\theta) = \frac{\left| \mathbf{a}^{H}_{K}(\boldsymbol\theta_s) \mathbf{a}_{K}(\boldsymbol\theta) \left[\left(\mathbf{d}(\boldsymbol\theta_s) \otimes \mathbf{b}(\boldsymbol\theta_s)\right)^{H} \left(\mathbf{d}(\boldsymbol\theta) \otimes \mathbf{b}(\boldsymbol\theta)\right)\right]\right|^2}{\left\|\mathbf{a}^{H}_{K}(\boldsymbol\theta_s)\right\|^4  \left\|\mathbf{d}(\boldsymbol\theta_s) \otimes \mathbf{b}(\boldsymbol\theta_s) \right\|^4  }
\end{equation} 

\section{Proposed Radar-Centric Spectrum Sharing}
\label{sec:NSP}
In this section, we introduce a radar-centric projection algorithm which projects the overlapped-MIMO radar signal onto the null space of the communication interference channel via the null space projection (NSP) technique proposed in~\cite{Awais_Spatial}. The algorithm requires us to have CSI in advance, which can be obtained in a number of ways and conveyed to the radar via mutual cooperation between the communication and radar systems~\cite{Shabnam2012GC, KhawarICNC14, Awais_Spatial}. 

The proposed algorithm works as follows. The radar receives $\mathbf{H}$, the CSI between radar and communication channels at the beginning. It then calculates the number of null spaces available to project, which is $M_T-N_R$. It then calculates the projection channel matrix $\mathbf{P}$ and finally creates a new radar signal $\mathbf{\hat{x}_{Radar}}$. If $\mathbf{H}$ is the channel matrix and $\mathbf{P}$ is the projection matrix onto the null space of $\mathbf{H}$, then the overlapped-MIMO radar waveform projected onto the null space of $\mathbf{H}$ to avoid interference from radar can be written as

\begin{equation}
\mathbf{\hat{x}_{Radar}} (t) = \mathbf{P} \mathbf{x_{Radar}} (t).
\end{equation} 

\subsection{Projection Matrix} 

Let channel $\mathbf H \in \mathbf{F}^{m \times n}$ for $\mathbf{F} = \mathbb{R}$ or $\mathbf{F} = \mathbb{C}$. We want a projection $\mathbf P \in \mathbf{F}^{n \times n}$ of a maximum rank such that it satisfies

\begin{itemize}
	\item $\mathbf{H}\mathbf{P} = 0$ 
	\item $\mathbf{P}^2 = \mathbf{P}$
\end{itemize}

Let the SVD of channel $\mathbf{H}$ be $\mathbf{H} = \mathbf{U} \boldsymbol\Sigma \mathbf{V}^{*}$, where $\mathbf{U}$ and $\mathbf{V}$ are unitary or orthogonal, depending on $\mathbf{F}$, of order $m$ and $n$, respectively, and 

\begin{description}
	\item[$\boldsymbol\Sigma = diag (\sigma_1, \sigma_2,\cdots, \sigma_k) \in \mathbb{R}^{m \times n},$ \: \: $k=\min{(m,n)}$] 
	\item[where $\sigma_{1} \geq \cdots \geq \sigma_{p} > \sigma_{p+1} = \cdots \sigma_{k} = 0.$]
\end{description}

Let $\boldsymbol\Sigma^{'} = diag (\sigma^{'}_1, \sigma^{'}_2,\cdots, \sigma^{'}_n) \in \mathbb{R}^{n \times n}$ such that

\[
 \sigma^{'}_i =
  \begin{cases}
   0 & \text{if } i \leq p \\
   1 & \text{if } i > p
  \end{cases}
\]

Note that $\boldsymbol\Sigma \boldsymbol\Sigma^{'} = 0$ and $\left(\boldsymbol\Sigma^{'}\right)^2 = \boldsymbol\Sigma^{'}$.

Now, we define projection $\mathbf{P} = \mathbf{V} \boldsymbol\Sigma^{'} \mathbf{V}^{*}$ and check the desired properties:

\begin{itemize}
  \item $\boldsymbol\Sigma \boldsymbol\Sigma^{'} = 0$ 
	\item $(\boldsymbol\Sigma^{'})^2 = \boldsymbol\Sigma^{'}$
	\item $\mathbf{H}\mathbf{P} = \mathbf{U} \boldsymbol\Sigma \mathbf{V}^{*} \mathbf{U} \boldsymbol\Sigma^{'} \mathbf{V}^{*} = 0$
	\item $\mathbf{P}^2 = \mathbf{V} \boldsymbol\Sigma^{'} \mathbf{V}^{*} \mathbf{V} \boldsymbol\Sigma^{'} \mathbf{V}^{*} = \mathbf{P}$ 
\end{itemize}

So $\mathbf{P}$ is the desired projection. 

\subsection{Optimal Subarray Size}

We run into two possible scenarios: (1) $M_T\leq N_R$ and (2) $M_T > N_R$. For first scenario where we have $M_T\leq N_R$, we cannot use the NSP method. However, a possible solution to this problem is using overlapped-MIMO as it increases the effective number of transmit arrays, thus making NSP possible. In this case, the effective transmit array aperture, $M_{\epsilon}$ is equal to $(M_T - K + 1)K$, which is greater than $N_R$. Note that $M_{\epsilon}$ is essentially the number of the virtual arrays in the transmitter of the radar. Hence, the overlapped-MIMO radar results in a total virtual array of size $\left( (M_T - K + 1)K \right) M_R$. On the other hand, if we have $M_T > N_R$, NSP is possible for $M_T-N_R$ dimensions. However, even in this case the performance can be increased using overlapped-MIMO since it increases the effective number of transmit arrays. 

In order to maximize the impact of the projection presented we have to select a value for the number of subarrays $K$ that maximizes $M_{\epsilon}$

\begin{equation}
K = \arg\max_{K} \left( M_{\epsilon} \right)
\end{equation}

\noindent where $M_{\epsilon} = \left(M_T - K + 1\right)K$.

The number of subarrays $K$ in the overlapped array can be optimized by 
\begin{eqnarray}
\label{S5EQxx}
\frac{\partial}{\partial K} \left(M_{\epsilon}\right) &=& 0 \\ \notag
\frac{\partial}{\partial K} \Big(\left(M_T - K + 1\right)K\Big) &=& 0 \\ \notag
                                              M_T - 2K + 1 &=& 0 \\ \notag
											                                   K &=&  \left\lfloor \frac{M_T + 1}{2} \right\rfloor \notag 
\end{eqnarray}

\noindent where $\left\lfloor \cdot \right\rfloor$ stands for the floor operation as $K$ should be integer.

\section{Simulation Results}
\label{sec:results}

We simulate an overlapped-MIMO radar. We assume a ULA with $M_T = 20$ antenna elements at the transmitter. At the receiver, we also assume $M_R=20$ antennas. In both cases the space between elements is $d_T=0.5$, meaning adjacent antenna elements are half a wavelength apart. The signal passes through a Rayleigh distributed channel and is subject to AWGN. Each antenna element is omnidirectional. We assume the target of interest is at $\theta_s=15^{o}$ and two interfering signals are located at directions $-30^{o}$ and $-10^{o}$. For the overlapped-MIMO radar we used $5$ subarrays, meaning $K=5$. Output SINRs are computed using $100$ independent simulations. 

\begin{figure}
\centering
\includegraphics[width=3.2in]{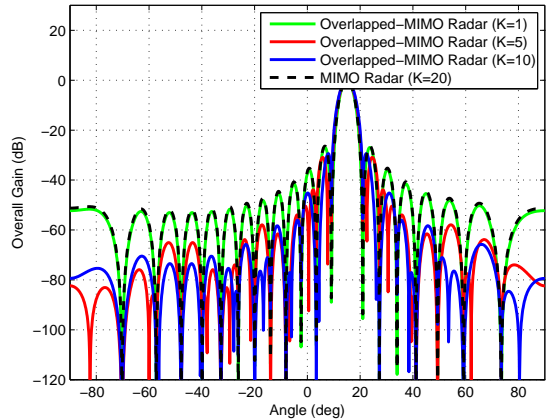}
\caption{Overall beampattern using conventional transmit-receive beamformer where the total number of elements is $M_T = 20$, the number of overlapped subarrays is $K = 5$ and $K = 10$ respectively, the number of elements in each subarray is $\left(M_T - K + 1\right) = 16$ and $\left(M_T - K + 1\right) = 11$ respectively, and $d_T = 0.5$ wavelength.}
\label{fig:omimo}
\end{figure}

Figure~\ref{fig:omimo} shows the overall beampattern for four different MIMO radar formulations: (1) overlapped-MIMO radar with $K =1$ (meaning $1$ subarray), (2) overlapped-MIMO radar with $K =5$, (3) overlapped-MIMO radar with $K =10$ and (4) MIMO radar with $K =20$ (pure MIMO). Here the overlapped-MIMO radars have two different orientations of $5$ and $10$ overlapped subarrays and each subarray has $11$ and $16$ antenna elements respectively. We can observe that the overlapped-MIMO with $K=1$ and MIMO radars have exactly the same overall transmit/receive beampatterns. However, the overlapped-MIMO radar has significantly improved sidelobe suppression compared to the beampattern of the pure MIMO radar (or $1$ subarray).

\begin{figure}
\centering
\includegraphics[width=3.2in]{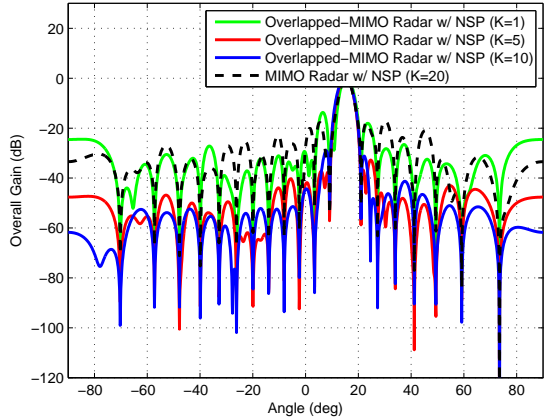}
\caption{Overall beampattern using conventional transmit-receive beamformer and NSP where the total number of elements is $M_T = 20$, the number of overlapped subarrays is $K = 5$ and $K = 10$ respectively, the number of elements in each subarray is $\left(M_T - K + 1\right) = 16$ and $\left(M_T - K + 1\right) = 11$ respectively, and $d_T = 0.5$ wavelength.}
\label{fig:omimonsp}
\end{figure}

Figure~\ref{fig:omimonsp} shows the overall beampattern for four different MIMO radar formulations with NSP algorithm: (1) overlapped-MIMO radar with $K =1$ plus NSP (meaning $1$ subarray), (2) overlapped-MIMO radar with $K =5$ plus NSP, (3) overlapped-MIMO radar with $K =10$ plus NSP and (4) MIMO radar with $K =20$ plus NSP (meaning pure MIMO). We observe that the projection algorithm has reduced sidelobe suppression as expected. Note that it is still providing encouraging suppression in compared to pure MIMO radar. However, the primary benefits are at the communications side since this NSP algorithm minimizes interference form the radar to the communications system and thus, enables the two to coexist. 

\begin{figure}
\centering
\includegraphics[width=3.2in]{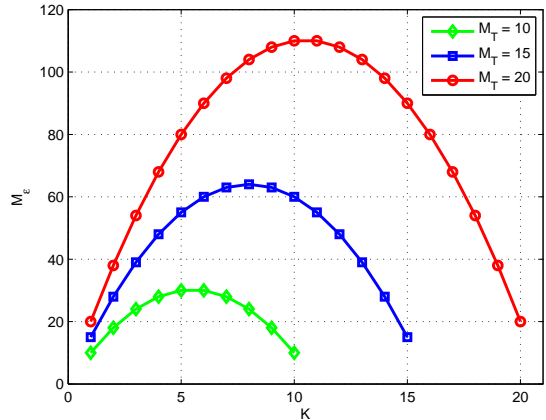}
\caption{The number of subarrays, $K$, in a overlapped-MIMO radar is varied from $1$ to $M_T$ and the resulting effective virtual transmitter array number, $M_{\epsilon}$ is observed for three different transmit antenna sizes, i.e., $M_T = 10$, $M_T = 15$ and $M_T = 20$.}
\label{fig:KMax}
\end{figure}

The final experiment considers the number of subarrays, $K$, in the transmitter of the overlapped-MIMO radar that maximizes the benefit for the radar in terms of sidelobe suppression. Note that the radar has most significant impact when the number of virtual arrays, $M_{\epsilon}$, on transmitter side is maximized (see equation~\ref{S5EQxx}). Fig.~\ref{fig:KMax} shows the impact of varying the number of subarrays $K$ from $1$ to $M_T$ on $M_{\epsilon}$. For $M_T = 20$, $K = 11$ or $K = 12$ results in the highest impact. This knowledge enables determining the structure of overlapping subarrays. The plot of $K$ for $M_T = 10$ and $M_T = 15$ are shown in the same figure to provide a comparative view. This graph enables picking a value for $K$ (the number of subarrays in the overlapped-MIMO structure) that maximizes the virtual antenna array size, thus enhancing the amount of sidelobe suppression in radar beampattern, while retaining the dimension needed for NSP.

\section{Conclusion}
\label{sec:conclusion}
This paper has explored an overlapped-MIMO antenna architecture and a spectrum sharing algorithm via null space projection (NSP) for radar-communications coexistence. In the overlapped-MIMO architecture, the transmit array of the radar is partitioned into a number of subarrays that are allowed to overlap. Each of the antenna elements in these subarrays have signals orthogonal to each other and to the elements of the other subarrays. The advantage of such architecture is to have a larger \textit{effective} transmit array with increased diversity gain. This formulation also improves overall sidelobe suppression compared to a conventional MIMO radar making it suitable for coexisting with communications system. 

Further, we introduced a radar-centric spectrum sharing algorithm that projects the radar signal onto the null space of the communications system's interference channel, which helps to avoid interference from the radar. Note that such null space projection is only possible when the physical number of transmit antennas of the radar is greater than the number of receive antennas of the communications system. 

Analytical models for the waveform of the overlapped-MIMO radar and the NSP algorithm are derived in this paper. Simulations of the coexistence scenario were presented too. Through analytical derivation and as well, simulation results, we were able to show that the proposed overlapped-MIMO and NSP algorithms outperform conventional schemes and enable radar-communications system coexistence in the same band. We found that the overlapped-MIMO architecture achieves more than $20$ dB sidelobe suppression above conventional MIMO radar when there are $20$ physical antenna elements in the radar system. We also observed that, in a similar setup, even though NSP degrades suppression, it still retains more than $10$ dB additional sidelobe suppression compared to conventional MIMO radar while reducing interference to the communications system.
%
\section*{Acknowledgment}
The authors would like to thank Awais Khawar for useful discussions on the projection-based spectrum sharing algorithm. 
\bibliography{IEEEabrv,HRadar}
\bibliographystyle{ieeetr}
%
\end{document}